\documentclass[12pt, draftclsnofoot, onecolumn]{IEEEtran}
\usepackage[cmex10]{amsmath}
\usepackage{amssymb}
\usepackage{cite}
\usepackage{graphicx}
\usepackage{array,color}
\usepackage{multirow}
\usepackage{amsmath,esint}
\allowdisplaybreaks[4]
\usepackage{lipsum}
\usepackage{cuted}
\usepackage{stfloats}
\usepackage{graphicx}
\usepackage{subfigure}
\usepackage{tabularx}
\usepackage{epsfig,epsf,color,balance,cite}
\usepackage{setspace}
\usepackage{bm}
\usepackage{textcomp}
\usepackage{algorithmic}
\usepackage{algorithm}
\usepackage{comment}
\usepackage{subfigure}
\usepackage{caption}
\usepackage{graphicx}
\usepackage{setspace}
\usepackage{diagbox}
\usepackage{float}
\begin{document}

\title{Quantized RIS-aided mmWave Massive MIMO Channel Estimation with Uniform Planar Arrays}

\author{Ruizhe Wang, Hong Ren, Cunhua Pan, Shi Jin, Petar Popovski, \IEEEmembership{Fellow, IEEE}, and Jiangzhou Wang, \IEEEmembership{Fellow, IEEE}
\thanks{ R. Wang, H. Ren, C. Pan and S. Jin are with National Mobile Communications Research Laboratory, Southeast University, Nanjing 210096, China. (e-mail:{rzwang, hren, cpan, jinshi}@seu.edu.cn).
Petar Popovski is with the Department of Electronic Systems, Aalborg University, 9220 Aalborg, Denmark (e-mail: petarp@es.aau.dk).
Jiangzhou Wang is with the School of Engineering, University of Kent, CT2 7NZ Canterbury, U.K. (e-mail: j.z.wang@kent.ac.uk).}
}

\maketitle

%\thanks{This work was supported in part by the National Key Research and Development Project under Grant 2019YFE0123600, National Natural Science Foundation of China (62101128), Basic Research Project of Jiangsu Provincial Department of Science and Technology (BK20210205) and High Level Personal Project of Jiangsu Province (JSSCBS20210105)}}

\begin{abstract}
\begin{comment}
	Millimeter wave (mmWave) massive multiple-input multiple-output (massive MIMO) is one of the promising technologies for the fifth-generation (5G) wireless communication systems.
\end{comment} 
In this paper, we investigate a cascaded channel estimation method for a millimeter wave (mmWave) massive multiple-input multiple-output (MIMO) system aided by a reconfigurable intelligent surface (RIS) with the BS equipped with low-resolution analog-to-digital converters (ADCs), where the BS and the RIS are both equipped with a uniform planar array (UPA). Due to the sparse property of mmWave channel, the channel estimation can be solved as a compressed sensing (CS) problem. However, the low-resolution quantization cause severe information loss of signals, and traditional CS algorithms are unable to work well. To recovery the signal and the sparse angular domain channel from quantization, we introduce Bayesian inference and efficient vector approximate message passing (VAMP) algorithm to solve the quantize output CS problem. To further improve the efficiency of the VAMP algorithm, a Fast Fourier Transform (FFT) based fast computation method is derived. Simulation results demonstrate the effectiveness and the accuracy of the proposed cascaded channel estimation method for the RIS-aided mmWave massive MIMO system with few-bit ADCs. Furthermore, the proposed channel estimation method can reach an acceptable performance gap between the low-resolution ADCs and the infinite ADCs for the low signal-to-noise ratio (SNR), which implies the applicability of few-bit ADCs in practice.
\end{abstract}


\IEEEpeerreviewmaketitle
%
%\newpage
\section{Introduction}
Millimeter wave (mmWave) massive multiple-input multiple-output (massive MIMO) can exploit an excessive number of spatial degrees of freedom by equipping multiple antennas \cite{mmwaveMIMO} and provide a large antenna array gain, which can significantly ameliorate the problem that the mmWave signal suffers from severe attenuation. Furthermore, compared with sub-6 GHz frequency band, the shorter wavelength of mmWave carrier means smaller size of the antenna array, which is more applicable for short range communication \cite{mmwaveMIMO2}.

However, the increase of power consumption and hardware cost cannot be ignored in mmWave massive MIMO systems. On the one hand, the converting rate of the analog-to-digital converters (ADCs) should increase accordingly as the bandwidth increases. On the other hand, research on ADC \cite{ADC1} showed that the power consumption of ADC will dramatically increase as the sampling rate grows up. Hence, the high-speed and high-precision ADCs are required to obtain accurate conversion of analog signal received at the antenna array into digital signal, which incurs high cost and power consumption. To tackle these problems, equipping high-speed but low-resolution ($\leq 4$ bits) ADCs for a large antenna array is a promising solution \cite{wenchaokai}. Many existing contributions have analyzed the throughput or uplink achievable rate when the BS is equipped with few-bit ADCs \cite{zkd,lf}, which clarified the feasibility of using low-resolution ADCs.\begin{comment}
	However, they either ignored the specific channel estimation process, or assumed that the channel state information (CSI) is well known, or simply using the linear minimum-mean-square-error (LMMSE) algorithm by modeling the quantization distortion as an additive noise.
\end{comment} 

\vspace{-0.0cm}
In addition, by reconfiguring the wireless propagation environment and providing an alternative propagation path for wireless signals, reconfigurable intelligent surfaces (RISs) are expected to be applied in mmWave systems to improve the spectrum efficiency and address the issue that the mmWave signals are vulnerable to blockages \cite{pengzd}. However, considering that the few-bit ADCs are employed at the BS instead of infinite-bit ADCs, the coarse quantization process can incur severe information loss. As the dimensions of the array of the BS and that of the RIS increases, the number of channel coefficients can be very large, and thus it is very difficult to reconstruct the cascaded channel from the coarse quantization process output.

In this paper, we consider the cascaded channel estimation scheme for an RIS-aided mmWave massive MIMO system with low-resolution ADCs. Due to the sparse property of mmWave channel, the channel estimation can be constructed as a sparse signal recovery problem, which can be solved using the compressed sensing (CS)-based algorithms. However, coarse quantization causes severe loss of information for signals, making the general CS algorithms no longer applicable. To tackle with this problem, we propose a vector approximate message passing (VAMP)-based cascaded channel estimation strategy. Specifically, the Bayesian inference and VAMP algorithm are adopted to recover the received signals and estimate the angular domain channel iteratively. Then, a Fast Fourier Transform (FFT)-based fast implementation is derived, which can dramatically reduce the computational complexity. Simulation results clarify the high efficiency and low complexity of the proposed channel estimation method.
\begin{comment}
At low and medium SNRs, there is only marginal performance loss between few-bit ADCs and ideal infinite-bit ADCs, which demonstrates the feasibility of applying low-precision ADCs in practice.
\end{comment}
\section{System Model}\label{systemmodel}
\begin{figure}[t]
\begin{minipage}[t]{0.99\linewidth}
\centering
\includegraphics[width=3.2in]{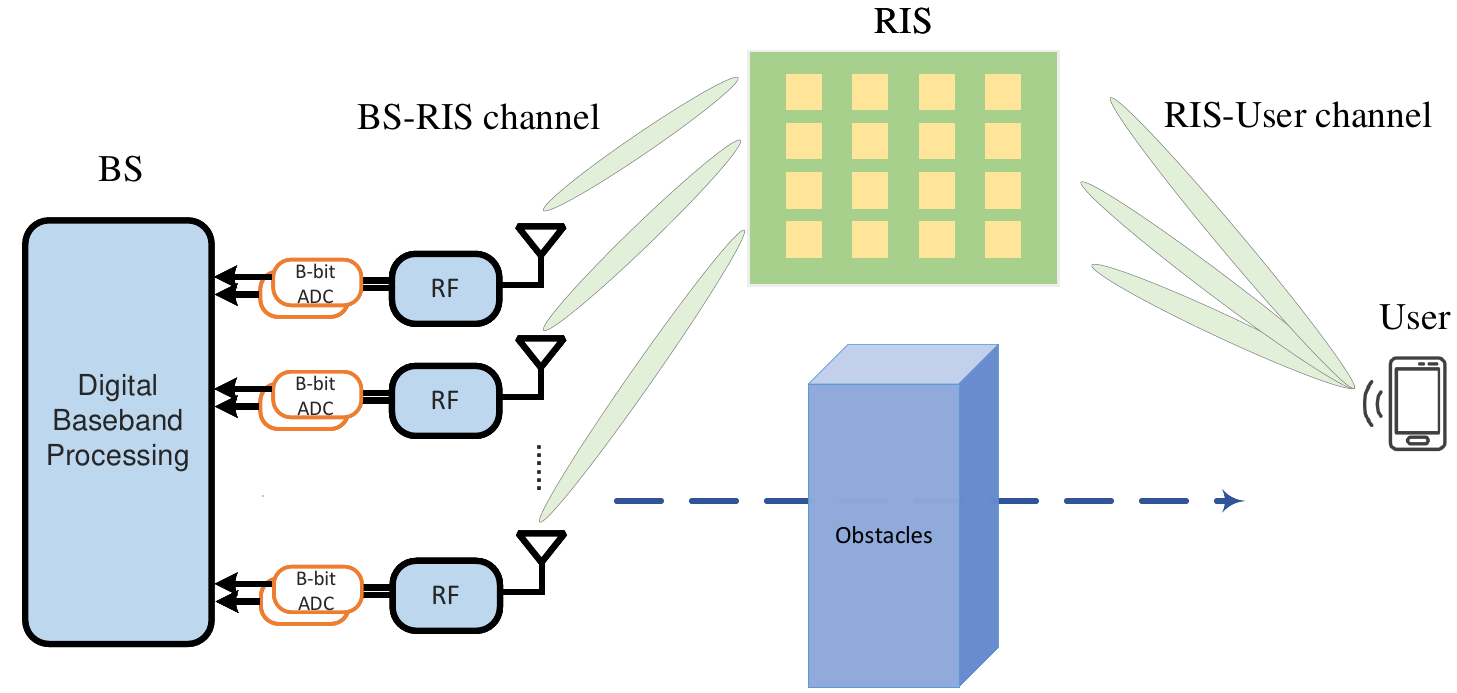}
\vspace{-0.2cm}
\caption{An RIS-aided uplink mmWave MIMO system where the BS is equipped with low-resolution ADCs.}
\label{fig0}\vspace{-0.7cm}
\end{minipage}%
\hfill
\end{figure}

Consider a narrow-band time-division duplex (TDD) mmWave massive MIMO system with low-resolution ADCs, as shown in Fig.~\ref{fig0}. The BS is equipped with an $N = N_1\times N_2$ antennas uniform planar array (UPA) to serve a multi-antenna user, where $N_1$ and $N_2$  are the numbers of antennas in the vertical dimension and the horizontal dimension, respectively. The user is equipped with a $Q = Q_1\times Q_2$ UPA. Meanwhile, a UPA-type RIS that is composed of $M = M_1\times M_2$ phase shift elements is deployed. In this paper, the quasi-static and block fading channel model is adopted, where the channels stay constant in each coherence block. Using the Saleh-Valenzuela (SV) model, the array response vectors of a
 $K = K_1\times K_2$ UPA can be denoted as
\begin{equation}\label{UPAarray}
{\bf a}_K(z,x) = {\bf a}_{K_1}(z)\otimes{\bf a}_{K_2}(x),
\end{equation}

\noindent where ${\bf a}_{K_1}(z)=[1,e^{-\text{j}2\pi z},\cdots,e^{-\text{j}(K_1-1)2\pi z}]^T$ is the array steering vector of the vertical direction of the UPA and ${\bf a}_{K_2}(x)=[1,e^{-\text{j}2\pi x},\cdots,e^{-\text{j}(K_2-1)2\pi x}]^T$ is the array steering vector of horizontal direction, and $\otimes$ denotes the Kronecker product. Denote $\vartheta\in [-90^\circ,90^\circ)$ and $\varrho\in [-180^\circ,180^\circ)$ as the zenith angle and azimuth angle of the UPA, respectively. The spatial frequency pair $(z,x)$ can be denoted by the physical angle pair $(\vartheta,\varrho)$ as \begin{equation}\label{angle}
z = \frac{d}{\lambda_{\text c}}\cos(\vartheta),\quad x = \frac{d}{\lambda_{\text c}}\sin(\vartheta)\cos(\varrho),
\end{equation}
where $d$ is the element spacing and $\lambda_{\text c}$ is the carrier wavelength. In this paper, we assume that $d_{\text{BS}} = d_{\text{RIS}} = \lambda_{\text c}/2$. Denote ${\bf H}_{\text{br}}\in \mathbb{C}^{N\times M}$ as the channel from the RIS to the BS, and ${\bf H}_{\text{ru}}\in \mathbb{C}^{M\times Q}$ as the channel from the user to the RIS. From (\ref{UPAarray}) and (\ref{angle}), the channel matrix ${\bf H}_{\text{br}}$ and ${\bf H}_{\text{ru}}$ can be expressed as
\begin{align}
{\bf H}_{\text{br}}&=\sum_{l=1}^L \alpha_l{\bf a}_N(\theta_l,\varphi_l){\bf a}^H_M(\phi_l,\upsilon_l),\label{Hbr}\\
{\bf H}_{\text{ru}}&=\sum_{j=1}^J{\beta_j{\bf a}_M(\omega_j,\psi_j){\bf a}^H_Q(\varpi_j,\kappa_j)},
\end{align}
where $L$ and $J$ represent the numbers of propagation paths between the BS and the RIS and between the RIS and the user, respectively. The parameters $\alpha_l$ and $\beta_j$ denote the complex path gain of the $l$-th path in the RIS-BS channel and that of the $j$-th path in user-RIS channel, respectively. The angles pair $(\theta_l,\varphi_l)$ and $(\phi_l,\upsilon_l)$ are the AoAs and AoDs of the $l$-th path in the RIS-BS channel, $(\omega_j,\psi_j)$ and $(\varpi_j,\kappa_j)$ represent the AoAs and AODs of the $j$-th path in the user-RIS channel, respectively. Assume that the RIS phase configuration changes $P$ times during the channel estimation process, and the user sends $T$ time slots of pilot signals in each RIS phase configuration. Then, the received signal at the $p$-th RIS phase configuration ${\bf e}^p$ at time slot $t$ and the sum of $T$ time slots are respectively given by
\begin{align}
	 \mathbf{y}_{t}^{p}&={{\mathbf{H}}_{\text{br}}}\text{Diag}\left( {{\mathbf{e}}^{p}} \right){{\mathbf{H}}_{\text{ru}}}{{\mathbf{s}}_{t}}+\mathbf{w}_{t}^{p}, \\
	 {{\mathbf{Y}}^{p}}&=\left[ \mathbf{y}_{1}^{p},\cdots ,\mathbf{y}_{T}^{p} \right]={{\mathbf{H}}_{\text{br}}}\text{Diag}\left( {{\mathbf{e}}^{p}} \right){{\mathbf{H}}_{\text{ru}}}\mathbf{S}+{{\mathbf{W}}^{p}},
\end{align}
where ${\bf s}_t, 1\leq t \leq T$ and ${\bf w}^p_t, 1\leq t \leq T$ represent the pilot signals transmitted in time slot $t$ and the received noise at the $p$-th RIS phase configuration in time slot $t$, respectively, and ${\bf S}=[{\bf s}_1,\cdots,{\bf s}_T]\in\mathbb{C}^{Q\times T}$,  ${\bf W}^p=[{\bf w}^p_1,\cdots,{\bf w}^p_T]\in\mathbb{C}^{N\times T}$. 
\vspace{-0.5cm}

\subsection{Cascaded Channel Sparsity Model}
By using the virtual angular domain representation, we can approximate the BS-RIS channel ${\bf H}_{\text{br}}$ and the RIS-user channel ${\bf H}_{\text{ru}}$ as
\begin{align}\label{sparsemodel}
	{\bf H}_{\text{br}}&\approx{\bf F}_N{\widetilde{\bf H}_{\text{br}}}{\bf F}^H_M,\nonumber\\
	{\bf H}_{\text{ru}}&\approx{\bf F}_M{\widetilde{\bf H}_{\text{ru}}}{\bf F}^H_Q,
\end{align}
where ${\widetilde{\bf H}_{\text{br}}}\in\mathbb{C}^{N\times M}$ and ${\widetilde{\bf H}_{\text{ru}}}\in\mathbb{C}^{M\times Q}$ are the angular cascaded channel of the BS-RIS channel and the RIS-user channel, respectively. ${\bf F}_N\in\mathbb{C}^{N\times N}$, ${\bf F}_M\in\mathbb{C}^{M\times M}$ and ${\bf F}_Q\in\mathbb{C}^{Q\times Q}$ are the dictionary matrices. With an $N_1\times N_2$ UPA at the BS, an $M_1\times M_2$ UPA at the RIS and a $Q_1\times Q_2$ UPA at the user, we have
\begin{align}
	{\bf H}_{\text{br}}&\approx({\bf U}_{N_1}\otimes {\bf U}_{N_2}){\widetilde{\bf H}_{\text{br}}}({\bf U}^H_{M_1}\otimes {\bf U}^H_{M_2}),\nonumber\\
	{\bf H}_{\text{ru}}&\approx({\bf U}_{M_1}\otimes {\bf U}_{M_2}){\widetilde{\bf H}_{\text{ru}}}({\bf U}^H_{Q_1}\otimes {\bf U}^H_{Q_2}),
\end{align}
where ${\bf U}_n\in \mathbb{C}^{n
	\times n}$ denotes the $n$-dimensional discrete Fourier transform (DFT) matrix. In our work, the angles are continuously and uniformly distributed and not necessarily located on grid, which is more practical.
By aggregating all signals ${\bf Y}^p$ in $P$ RIS phase configuration, the received signal matrix is given by
\begin{align}\label{MulRecv}
	\mathbf{Y}&=\left[ \text{vec}\left( {{\mathbf{Y}}^{1}} \right),\cdots ,\text{vec}\left( {{\mathbf{Y}}^{P}} \right) \right] + \left[\text{vec}({\bf W}^1),\cdots, \text{vec}({\bf W}^P)\right]\nonumber\\
	&\overset{a}{=}{{\left( {{\mathbf{H}}_{\text{ru}}}\mathbf{S} \right)}^{T}}\diamond {{\mathbf{H}}_{\text{br}}}\mathbf{E}+ {\bf W} \nonumber\\
	&={{{\mathbf{S}^{T}{\mathbf{H}}^{T}_{\text{ru}}}}}\diamond {{\mathbf{H}}_{\text{br}}}\mathbf{E}+ {\bf W} \\
	& ={{\left( {{\mathbf{F}}_{M}}{{{\mathbf{\widetilde{H}}}}_{\text{ru}}}\mathbf{F}_{Q}^{H}\mathbf{S} \right)}^{T}}\diamond \left( {{\mathbf{F}}_{N}}{{{\mathbf{\widetilde{H}}}}_{\text{br}}}\mathbf{F}_{M}^{H} \right)\mathbf{E}+ {\bf W} \nonumber\\
	& \overset{b}{=}\left( {{\mathbf{S}}^{T}}\mathbf{F}_{Q}^{*}\otimes {{\mathbf{F}}_{N}} \right)\left( \mathbf{\widetilde{H}}_{\text{ru}}^{T}\otimes {{{\mathbf{\widetilde{H}}}}_{\text{br}}} \right)\left( \mathbf{F}_{M}^{T}\diamond \mathbf{F}_{M}^{H} \right)\mathbf{E}+ {\bf W},\nonumber
\end{align}
where ${\bf E}=\left[{\bf e}^1,\cdots,{\bf e}^P\right]$, and the equality $a$ has used the property that $\text{vec}({\bf A}\text{Diag}({\bf b}){\bf C}) = ({\bf C}^T\diamond{\bf A}){\bf b}$ \cite{tensor}, and the operator $\diamond$ denotes the Khatri-Rao product. The equality $b$ is obtained via the property that $({\bf A}\otimes {\bf B})({\bf C}\diamond {\bf D})={\bf AC}\diamond{\bf BD}$. The matrices ${{\mathbf{\widetilde{H}}}}_{\text{br}}$ and  ${{\mathbf{\widetilde{H}}}}_{\text{ru}}$ represent the sparse angular domain matrix of the RIS-BS channel and the user-RIS channel, respectively. We aim to estimate the cascaded channel ${\mathbf{H}}^T_{\text{ru}}\diamond {\mathbf{H}}_{\text{br}}$. From (\ref{MulRecv}), it can be seen that the sparse matrix of the RIS-aided multi-antenna user cascaded channel $\mathbf{\widetilde{H}}_{\text{ru}}^{T}\otimes {{{\mathbf{\widetilde{H}}}}_{\text{br}}} $ is expressed in the form of the Kronecker product, leading to the substantial increase in the number of parameters that needs to be estimated. Fortunately, Proposition 1 in \cite{junfang} showed that the Khatri-Rao product $\mathbf{F}_{M}^{T}\diamond \mathbf{F}_{M}^{H}$ only contains $M$ distinct rows, which means that the other $M(M-1)$ rows still belong to the first $M$ rows. Based on the above discussions, the cascaded channel matrix can be further compressed, and (\ref{MulRecv}) can be equivalently modified as
\begin{equation}\label{multiY}
	{\bf Y}=\underbrace{\left( {{\mathbf{S}}^{T}}\mathbf{F}_{Q}^{*}\otimes {{\mathbf{F}}_{N}} \right){\bf \Lambda}\mathbf{F}_{M}^{H}\mathbf{E}}_{\triangleq{\bf Z}}+{\bf W},
\end{equation}
where the sparse matrix ${\bf \Lambda}\in\mathbb{C}^{QN\times M}$ is a compressed version of $\left( \mathbf{\widetilde{H}}_{\text{ru}}^{T}\otimes {{{\mathbf{\widetilde{H}}}}_{\text{br}}} \right)\in\mathbb{C}^{QN\times M^2}$. More details about the proposition can be found in appendix. Then, the quantization output is given by
\begin{align}\label{multiQY}
	\widetilde{\bf Y}&=\mathcal{Q}\left({\bf Z + W}\right),
\end{align}
where $\mathcal{Q}$ denotes the quantize operation. 

\subsection{ADC Quantization Model}
Assume that each quantizer at the BS adopts uniform mid-rise quantization and the variable gain amplifiers are adopted. Denote $\Delta_{\text{Re}}$ and $\Delta_{\text{Im}}$ as the stepsizes of the real part quantizer and the imaginary part quatizer, respectively. Assuming that the input scalar $\xi$ is circularly symmetric, then $\Delta_{\text{Re}}=\Delta_{\text{Im}}=\sqrt{\mathbb{E}\{{|\xi|^2}/{2}\}}\Delta$. The coefficient $\Delta$ can be found in \cite{mojianhua}. For a $B$-bit quantizer, the analog input magnitude is divided into $2^B-1$ thresholds, which is denoted as $h_1, h_2, \cdots, h_{2^B-1}$, where $h_b = (-2^{B-1}+b)\Delta_{\text{Re}}, 1\leq b\leq2^B$.  Once the input signal falls in the interval $[h_{b-1}, h_{b})$, the output is set to $h_b - \frac{\Delta_{\text{Re}}}{2}$.\begin{comment}
	In general, the quantized output can be denoted as
	\begin{align}\label{quantize}
		\tilde{y}&=\mathcal{Q}(\xi)\nonumber\\
		&=\left(\text{min}\left(\text{max}\left(\left\lceil \frac{\text{Re}(\xi)}{\Delta_{\text{Re}}} \right\rceil, -2^{B-1}\right), 2^{B-1}-1 \right)-\frac{1}{2}\right)\Delta_{\text{Re}}\nonumber\\
		+&\text{j} \ \left(\text{min}\left(\text{max}\left(\left\lceil \frac{\text{Im}(\xi)}{\Delta_{\text{Im}}} \right\rceil, -2^{B-1}\right), 2^{B-1}-1 \right)-\frac{1}{2}\right)\Delta_{\text{Im}}.
	\end{align}
\end{comment}

\begin{comment}
For the case of one-bit ADCs, the quantized output is given by
\begin{align}
\tilde{y}&={\text{sign}}\left({\text{Re}}(\xi)\right)\sqrt{\frac{2}{\pi}}\left({\mathbb{E}\{{|\xi|^2}/{2}\}}\right)^{\frac{1}{2}}\Delta\nonumber\\
&\qquad+\text{j}\ {\text{sign}}\left({\text{Im}}(\xi)\right)\sqrt{\frac{2}{\pi}}\left({\mathbb{E}\{{|\xi|^2}/{2}\}}\right)^{\frac{1}{2}}\Delta.
\end{align}
\end{comment}

\begin{comment}
In practice, the stepsize of the ADC is controlled by a variable gain amplifier (VGA). Suppose that the average power $\mathbb{E}\{|\xi|^2\}$ has been measured in a long term before channel estimation and thus the stepsize of ADCs is already controlled. In addition, the uniform quantization is considered in this paper, but the nonuniform quantization (i.e., the adaptive quantization) is also applicable in our proposed method, which may bring better gains.
\end{comment}

\section{Channel Estimation Algorithm}\label{channelestimationalgorithm}
In this section, we aim to estimate the cascaded channel based on the quantized output $\widetilde{\bf Y}$ and the predetermined training matrix ${\bf E}$. The VAMP-based channel estimation strategy for a multi-antenna user is first described. Then, we propose the corresponding fast computation implementation for the channel estimation algorithm. Finally, the computational complexity of the fast implementation method is analyzed.
\vspace{-0.3cm}

\subsection{Problem Formulation}
Based on the above discussions, to estimate the cascaded channel, we only need to estimate the sparse matrix $\bf \Lambda$. Defining $\widetilde{\bf y} \triangleq \text{vec}\left(\widetilde{\bf Y}\right)$, ${\bf x}\triangleq \text{vec}\left( {\mathbf{{\Lambda }}} \right)$ and ${\bf w} \triangleq \text{vec}\left({\bf W}\right)$,  (\ref{multiQY}) can be vectorized as
\begin{align}\label{sparsequanti}
	\widetilde{\bf y}=\mathcal{Q}\left(\underbrace{{{\left( {{\mathbf{F}}^H_{M}}\mathbf{E} \right)}^{T}}\otimes \left({\bf S}^T \mathbf{F}_{Q}^{*}\otimes {{\mathbf{F}}_{N}} \right)}_{\triangleq{\bf A}}{\bf x} +{\bf w}\right).
\end{align}
From (\ref{sparsequanti}), the cascaded channel estimation can be formulated as the following problem: estimate the sparse vector ${\bf{x}}\in \mathbb{C}^{NMQ}$ from the observation $\tilde{\bf y}\in \mathbb{C}^{PTN} $ based on the known linear transform ${\bf{A}}\in \mathbb{C}^{P TN\times NMQ}$.
\begin{comment}
	From  (\ref{AHAM}), we can obtain the right singular matrix ${\bf V}_A\triangleq\frac{1}{\sqrt{MQ}}\left({\bf F}^T_M{\bf U}_E\otimes{\bf F}^T_Q{\bf U}_S\right)\otimes {\bf I}_N$, which satisfies ${\bf V}_A{\bf V}^H_A={\bf I}_{NMQ}$, and ${\bf S}^H_A{\bf S}_A=NMQ\left({\bf \Lambda}_E\otimes{\bf \Lambda}_S\right)\otimes{\bf I}_N$.
\end{comment}
\vspace{-0.3cm}

\subsection{VAMP Algorithm}
According to (\ref{sparsequanti}), the cascaded channel estimation can be formulated as a noisy quantized CS problem. To reconstruct the noiseless signal $\bf z$ and the sparse vector $\bf x$ from $\tilde{\bf y}$, we introduce the efficient VAMP algorithm \cite{VAMP} to implement the Bayesian optimal estimators. Assume that $\bf z$ is produced by a linear transform ${\bf z}={\bf Ax}$ by a sparse vector ${\bf x}$ with i.i.d elements and a known matrix ${\bf A}$. Based on the observation $\tilde{\bf y}=\mathcal{Q}({\bf Ax}+{\bf w})$ and the likelihood $p(\tilde{\bf y}|{\bf z})=\prod_i{p_{Y|Z}(y_i|z_i)}$, the VAMP algorithm can approach nearly minimum mean square error (MMSE) estimation of ${\bf x}$ by alternatively calculating the scalar estimation and linear MMSE (LMMSE) estimation iteratively. 

Without loss of generality, suppose the elements $x_i,\forall i$ follow the Bernoulli Gaussian-mixtrue (GM) with parameters $\bf q$ as
\begin{equation}\label{GM}
p_X\left(x_i ; {\bf q}\right)=\lambda_0 \delta\left(x_i\right)+\sum_i \lambda_i \mathcal{C N}\left(x_i ; \mu_i, \rho_i\right), \forall i,
\end{equation}
where ${\bf q}\triangleq\{\{\lambda_i\},\{\mu_i\},\{\rho_i\}\}, \forall i$ denotes the prior parameters set and $\lambda_0=\text{Prob}\{x=0\}$, and $\{\lambda_i\}$, $\{\mu_i\}$ and $\{\rho_i\}$ are the weights of the distributions, means and variances of the Gaussian mixture, respectively. 

Algorithm \ref{alg:1} gives the details of the VAMP algorithm, where $T_{\text{max}}$ is the number of maximum iteration time and it is set to 50 in our simulation. Steps 4 and 5 calculate the posterior mean and variance of $x$ based on the pseudo-prior $p_X(x)$ and the likelihood $p(x|r_1)$, where the quality ${\bf r}_1$ is the additive white Gaussian noise (AWGN) corrupted version of ${\bf x}$, i.e. ${\bf r}_k={\bf x}+\mathcal{CN}({\bf 0},\nu^x{\bf I})$, and the variance $\nu^x$ is computed iteratively. The posterior probability $p(x_{1m}|r_{1m})$ is given by
\begin{equation}
	p_{X|R}\left(x_m|\hat{r}_m; \nu_m^r, {\bf q}\right) \triangleq \frac{p_X\left(x_m ; {\bf q}\right) \mathcal{CN}\left(x_m ; \hat{r}_m, \nu_m^r\right)}{\int_x p_X(x ; {\bf q}) \mathcal{CN}\left(x ; \hat{r}_m, \nu_m^r\right)}.
\end{equation}
Step 6 updates the quality ${\bf r}_2$ by using an ``Osanger'' term $\alpha_k{\bf r}_1$. Steps 7 and 8 calculate the posterior mean and variance of $\bf z$ from the quantization output $\tilde{\bf y}=\mathcal{Q}({\bf z}+{\bf w})$. Assume that the $n$-th element of ${\bf z}$ follows the complex Gaussian distribution of $p(z_n) = \mathcal{CN}(z_n;\hat{p_n},\nu_p)$, then the posterior can be computed as 
\begin{equation}
	p_{Z|Y,P}\left(z_n | \tilde{y}_n ; \hat{p}_n, \nu_p\right) \triangleq \frac{p_{Y \mid Z}\left(\tilde{y}_n | z_n\right) \mathcal{CN}\left(z_n ; \hat{p}_n, \nu_p\right)}{\int_z p_{Y \mid Z}\left(\tilde{y}_n|z\right) \mathcal{CN}\left(z ; \hat{p}_n, \nu_p\right)}.
\end{equation}
The details about the derivations of posterior function $p(z|\tilde{y})$ and the posterior mean $\hat{z}$ and variance $\nu^z$ can be found in \cite{wenchaokai}. Then, based on the pseudo-prior of $z$ and $x$, steps 10 to 14 calculate the LMMSE estimation of ${\bf x}$ and ${\bf z}$ jointly.

\begin{algorithm}[h]
	\caption{VAMP-based Cascaded Channel Estimation Algorithm}
	\label{alg:1}
	\begin{algorithmic}[1]
		
		\REQUIRE $\widetilde{\bf Y}$, ${\bf p}_1$, ${\bf r}_1$, $p(\widetilde{\bf Y}|{\bf Z})$ and ${\bf q}$
		\ENSURE $\widehat{\bf x}$
		\STATE Initialize: $\hat{s}=0, \hat{x}_1=\int_x x p_X(x), \forall i,$\\$\nu^x_m=\int_x\left|x-\hat{x}_m(1)\right|^2 p_X(x)$
		\FOR{$t =1,\cdots,T_{\text{max}}$}
		\STATE $ \widehat{x}_{1 i} \leftarrow \mathbb{E}_{X \mid R}\left[x_i \mid r_{1 i} ; \nu^x_1, {\bf q}\right], \quad \forall i $\\
		\STATE $\alpha_1 \leftarrow  \frac{1}{\nu^x_1NMQ} \sum_{i=1}^{NMQ} \operatorname{Var}_{X \mid R}\left[x_i \mid r_{1 i} ; \nu^x_1, {\bf q}\right]$ \\
		\STATE $ \mathbf{r}_2 \leftarrow\left(\widehat{\mathbf{x}}_1-\alpha_1 \mathbf{r}_1\right) /\left(1-\alpha_1\right), \ \ \nu^x_2 \leftarrow \nu^x_1 \alpha_1 /\left(1-\alpha_1\right)$ \\
		\STATE $\widehat{z}_{1 i} \leftarrow \mathbb{E}_{Z \mid Y, P}\left[z_i \mid y_i, p_{1 i} ; \nu^p\right], \ \ \forall i $\\
		\STATE $\beta_1 \leftarrow \frac{1}{\nu^p_1 PNQ} \sum_{i=1}^{PNQ} \operatorname{Var}_{Z \mid Y, P}\left[z_i \mid y_i, p_{1 i} ; \nu^p_1\right] $\\
		\STATE $\mathbf{p}_2 \leftarrow\left(\widehat{\mathbf{z}}_1-\beta_1 \mathbf{p}_1\right) /\left(1-\beta_1\right), \quad \nu^p_2 \leftarrow \nu^p_1 \beta_1 /\left(1-\beta_1\right)$
		\STATE $\widehat{\mathbf{x}}_2 \leftarrow \mathbf{V}_A\left({\bf S}^H_A{\bf S}_A \nu^x_2 / \nu^p_2+\mathbf{I}\right)^{-1}\times$ \\
		$\left({\bf V}_A{\bf A}^H \mathbf{p}_2 \nu^x_2 / \nu^p_2\right. \left.+\mathbf{V}_A^H  \mathbf{r}_2\right)$, where  ${\bf V}_A$ and ${\bf S}^H_A{\bf S}_A$ are defined in (\ref{VA}), and each terms of $\widehat{\bf x}_2$ can be calculated via FFT similar to step 13\\
		\STATE $ \alpha_2 \leftarrow \frac{1}{NMQ} \sum_{n=1}^{NMQ} \nu^p_2 /\left(s_n^2 \nu^x_2+\nu^p_2\right) $\\
		\STATE $\mathbf{r}_1 \leftarrow\left(\widehat{\mathbf{x}}_2-\alpha_2 \mathbf{r}_2\right) /\left(1-\alpha_2\right), \ \ \nu^x_1 \leftarrow \nu^x_2 \alpha_2 /\left(1-\alpha_2\right)$
		\STATE Calculate $\widehat{\mathbf{z}}_2\leftarrow \mathbf{A} \widehat{\mathbf{x}}_2 $ by using FFT and (A1)-(A4)
		\STATE $\beta_2 \leftarrow\left(1-\alpha_2\right) PT/M $\\
		\STATE $\mathbf{p}_1 \leftarrow\left(\widehat{\mathbf{z}}_2-\beta_2 \mathbf{p}_2\right) /\left(1-\beta_2\right), \ \ \nu^p_1 \leftarrow \nu^p_2 \beta_2 /\left(1-\beta_2\right)$
		\ENDFOR
	\end{algorithmic}
\end{algorithm}

\vspace{-0.5cm}
\subsection{Fast Computation Implementation}
To tackle the information loss caused by few-bit quantization, the training length needs to reach hundreds to provide more measurements to estimate the channel. For the VAMP algorithm, most of the complexity comes from calculating the matrix-vector product ${\bf A}\widehat{\bf x}_2$ and ${\bf V}^H_A{\bf r}_2$ in each iteration, where , whose computational complexity is $\mathcal{O}(PN^2M)$ and $\mathcal{O}(N^2M^2)$, respectively.

To lower the computational complexity, a fast implementation method to compute ${\bf A}\widehat{\bf x}$ is proposed based on FFT. The received signal in (\ref{multiY}) can be further rewritten as
\begin{equation}
	{\bf Y}=\left( {{\mathbf{S}}^{T}}\mathbf{F}_{Q}^{*}\otimes {{\mathbf{F}}_{N}} \right)\left(\mathbf{E}^T\mathbf{F}_{M}^{\ast}{\bf \Lambda}^T\right)^T+{\bf W}.
\end{equation}
We can respectively  obtain $\mathbf{F}_{M}^{\ast}{\bf \Lambda}^T$ by performing 2-D inverse-FFT. Then, left multiple ${\bf E}^T$ by using fast convolution and obtain $\mathbf{E}^T\mathbf{F}_{M}^{\ast}{\bf \Lambda}^T$.

The computation of $\left( {{\mathbf{S}}^{T}}\mathbf{F}_{Q}^{*}\otimes {{\mathbf{F}}_{N}} \right)\left(\mathbf{E}^T\mathbf{F}_{M}^{\ast}{\bf \Lambda}^T\right)^T$ can be performed as follows. Denote ${\bf X}\triangleq\left(\mathbf{E}^T\mathbf{F}_{M}^{\ast}{\bf \Lambda}^T\right)^T\in \mathbb{C}^{NQ\times P}$ and reshape $\bf X$ as a three way tensor ${\mathcal{X}}\in \mathbb{C}^{N\times Q\times P}$. Then, the multiplication $\left( {{\mathbf{S}}^{T}}\mathbf{F}_{Q}^{*}\otimes {{\mathbf{F}}_{N}} \right){\bf X}$ can be rewritten as
\begin{align}\label{tensor1}
	\left( {{\mathbf{S}}^{T}}\mathbf{F}_{Q}^{*}\otimes {{\mathbf{F}}_{N}} \right){\bf X}=\left[ \text{vec}\left({\bf F}_N{\mathcal{X}}(:,:,1){\bf F}^{\ast}_Q{\bf S}\right),\cdots,\text{vec}\left({\bf F}_N{\mathcal{X}}(:,:,P){\bf F}^{\ast}_Q{\bf S}\right) \right].
\end{align}
Substituting ${{\mathbf{F}}_{N}}={{\mathbf{U}}_{{{N}_{1}}}}\otimes {{\mathbf{U}}_{{{N}_{2}}}}$ and ${{\mathbf{F}}_{Q}}={{\mathbf{U}}_{{{Q}_{1}}}}\otimes {{\mathbf{U}}_{{{Q}_{2}}}}$ into (\ref{tensor1}), we have
\begin{align}\label{tensor2}
	{{\mathbf{F}}_{N}}{\mathcal{X}}\left( :,:,p \right)\mathbf{F}_{Q}^{*}{\bf S}&=\left( {{\mathbf{U}}_{{{N}_{1}}}}\otimes {{\mathbf{U}}_{{{N}_{2}}}} \right){\mathcal{X}}\left( :,:,p \right)\left( \mathbf{U}_{{{Q}_{1}}}^{*}\otimes \mathbf{U}_{{{Q}_{2}}}^{*} \right){\bf S} \nonumber\\
	&=\left( {{\mathbf{U}}_{{{N}_{1}}}}\otimes {{\mathbf{U}}_{{{N}_{2}}}} \right){{\left({\bf S}^T \left( \mathbf{U}_{{{Q}_{1}}}^{*}\otimes \mathbf{U}_{{{Q}_{2}}}^{*} \right)({{{\mathcal{X} }}}\left( :,:,p \right))^{T} \right)}^{T}},\forall p.
\end{align}
Note that the $i$-th vector of the multiplication $\left( \mathbf{U}_{{{X}_{1}}}\otimes \mathbf{U}_{{{X}_{2}}} \right){\bf X}$ can be denoted as $\left( \mathbf{U}_{{{X}_{1}}}\otimes \mathbf{U}_{{{X}_{2}}} \right){\bf X}_{:,i} = {\text{vec}}(\mathbf{U}_{{{X}_{2}}}{\bf X}_i\mathbf{U}_{{{X}_{1}}}^T)$, where ${\bf X}_i = \text{unvec}({\bf X}_{:,i})$, and thus we can calculate each vector of the multiplication $\left( \mathbf{U}_{{{X}_{1}}}\otimes \mathbf{U}_{{{X}_{2}}} \right){\bf X}$ by using 2-D FFT. Similarly, we can calculate each vector of the multiplication $\left( \mathbf{U}_{{{X}_{1}}}^{*}\otimes \mathbf{U}_{{{X}_{2}}}^{*} \right){\bf X}$ by using 2-D IFFT. Based on the above discussions, the steps for fast computation of ${{\mathbf{F}}_{N}}{\mathcal{X}}\left( :,:,p \right)\mathbf{F}_{Q}^{*}{\bf S} $ are as follows:

\noindent(A1) Permute ${\mathcal{X}}\in \mathbb{C}^{N\times Q\times P}$ to ${\mathcal{X}}_1\in \mathbb{C}^{Q\times N\times P}$, and reshape it to ${\mathcal{X}}_2\in \mathbb{C}^{Q_1\times Q_2\times NP}$.

\noindent(A2)  Do 2-D inverse-FFT for each slice ${\mathcal{X}}_2(:,:,i)$, and then reshape to ${\mathcal{X}}_3\in \mathbb{C}^{Q\times N\times P}$.

\noindent(A3)  Calculate ${\bf S}^T{\mathcal{X}}_3(:,:,i)$, and permute to ${\mathcal{X}}_4\in\mathbb{C}^{N\times Q\times P}$, then reshape it to ${\mathcal{X}}_5\in\mathbb{C}^{N_1\times N_2\times QP}$.

\noindent(A4)  Do 2-D FFT for each slice ${\mathcal{X}}_5(:,:,i)$, and then reshape to $\widetilde{\bf X}\in\mathbb{C}^{NQ\times P}$, and we obtain $\widetilde{\bf X}=\left( {{\mathbf{S}}^{T}}\mathbf{F}_{Q}^{*}\otimes {{\mathbf{F}}_{N}} \right){\bf X}$.

\begin{comment}
	Based on the above discussions, the steps for fast computation of ${{\mathbf{F}}_{N}}{\mathcal{X}}\left( :,:,p \right)\mathbf{F}_{Q}^{*}{\bf S} $ are as follows.
	\begin{enumerate}
		\item Permute ${\mathcal{X}}\in \mathbb{C}^{N\times Q\times P}$ to ${\mathcal{X}}_1\in \mathbb{C}^{Q\times N\times P}$, and reshape it to ${\mathcal{X}}_2\in \mathbb{C}^{Q_1\times Q_2\times NP}$.
		\item Do 2-D inverse-FFT for each slice ${\mathcal{X}}_2(:,:,i)$, and then reshape to ${\mathcal{X}}_3\in \mathbb{C}^{Q\times N\times P}$.
		\item Calculate ${\bf S}^T{\mathcal{X}}_3(:,:,i)$, and permute to ${\mathcal{X}}_4\in\mathbb{C}^{N\times Q\times P}$, then reshape it to ${\mathcal{X}}_5\in\mathbb{C}^{N_1\times N_2\times QP}$.
		\item Do 2-D FFT for each slice ${\mathcal{X}}_5(:,:,i)$, and then reshape to $\widetilde{\bf X}\in\mathbb{C}^{NQ\times P}$, and we obtain $\widetilde{\bf X}=\left( {{\mathbf{S}}^{T}}\mathbf{F}_{Q}^{*}\otimes {{\mathbf{F}}_{N}} \right){\bf X}$.
	\end{enumerate}
	To further improve the efficiency, it is recommended that $T=Q$ and ${\bf S}={\bf I}_Q$ or ${\bf S}={\bf F}_Q$, which eliminates the need to calculate the multiplication ${\bf S}^T{\mathcal{X}}_3(:,:,i)$.
\end{comment}
The computation of the multiplication ${\bf V}^H_A {\bf r}$ follows a similar procedure. Specifically, let ${\bf S}={\bf I}_Q$ and we have
\begin{align}\label{VA}
	{\bf A}^H{\bf A}&=\left({\bf F}^T_M{\bf E}^\ast{\bf E}^T{\bf F}^\ast_M\right)\otimes NQ{\bf I}_{NQ}\nonumber\\
	&=NMQ\left(\frac{1}{\sqrt{M}}{\bf F}^T_M{\bf U}_E\otimes{\bf I}_{NQ}\right)\left({\bf \Lambda}_E\otimes{\bf I}_{NQ}\right)\left(\frac{1}{\sqrt{M}}{\bf U}^H_E{\bf F}^\ast_M\otimes{\bf I}_{NQ}\right)\nonumber\\
	&={\bf V}_A{\bf S}_A^H{\bf S}_A{\bf V}^H_A,
\end{align}
where ${\bf V}_A\triangleq\left(\frac{1}{\sqrt{M}}{\bf F}^T_M{\bf U}_E\otimes{\bf I}_{NQ}\right)$ and ${\bf U}_E$ and ${\bf \Lambda}_E$ are the eigen vectors and the eigenvalues of ${\bf E}^\ast{\bf E}^T$, respectively. Thus, the multiplication ${\bf V}^H_A{\bf r}$ can be also fast computed by using FFT similarly.

\subsection{Complexity Analysis}
For the VAMP algorithm, most of the computational complexity is introduced by the matrix multiplications ${\bf A}\widehat{\bf x}$. For a general measurement matrix $\bf A$, it needs the complexity of $\mathcal{O}(PMN^2Q^2)$. In our proposed algorithm, the fast computations of ${\bf E}^T{\bf F}^\ast_M{\bf \Lambda}^T$ need $NQ$ times of $P$ points FFT and IFFT, and $NQ$ times of $P$ points Hadamard products, the corresponding complexity is $\mathcal{O}(2NPQ\log P)$. Furthermore, the steps for fast computation of ${\bf F}_N{\mathcal{X}}(:,:,p){\bf F}^\ast_Q{\bf S}$ needs $NP$ times 2-D IFFT, $P$ times of matrix multiplication ${\bf S}^T{\mathcal{X}}(:,:,i)$ and $QP$ times of 2-D FFT. However, the signal matrix ${\bf S}$ can be simply substituted by ${\bf I}_Q$ or $Q$ points DFT matrix, and the matrix multiplication can be avoided, and the corresponding complexity is $\mathcal{O}(NPQ\log NQ)$. Hence, the total computational complexity for the VAMP algorithm is $T_{\text{max}}\mathcal{O}(2QPN\log QPN)$.

\section{Simulation Results}\label{sim}
In this section, the simulation results are presented to show the efficiency of the proposed method. Both the BS and the RIS are equipped with a UPA. The antenna spacing is half-wavelength. The SNR is defined as SNR $\triangleq\frac{\mathbb{E}\left[\|{\bf Z}\|^2_F\right]}{\mathbb{E}\left\{\|{\bf W}\|^2_F\right\}}$, where $\mathbb{E}\left\{\|{\bf Z}\|^2_F\right\}=\mathbb{E}\left\{\|({\bf S}^T{\bf H}^T_{\text{ru}}\diamond{\bf H}_{\text{br}}){\bf E}\|^2_F\right\}$. The scales of the UPAs employed at the BS and the RIS are $N_1=N_2=8$ and $M_1=M_2=8$, respectively. The multi-antenna user is equipped with a $2\times 2$ UPA, i.e., $Q_1=Q_2=2$. The set of angular parameters are generated from uniform distribution. The normalized mean square error (NMSE) is defined as NMSE$({\bf G})\triangleq\mathbb{E}\left\{\frac{\|{\bf G}-\widehat{\bf G}\|^2_F}{\|{\bf G}\|^2_F}\right\}$, where ${\bf G}\triangleq{\bf H}^T_{\text{ru}}\diamond{\bf H}_{\text{br}}$, and $\widehat{\bf G}$ is the estimated cascaded channel. The LS algorithm, the Bussgang theorem based LMMSE (BLM) algorithm, the quantized iterative hard thresholding (QIHT) algorithm \cite{qiht}, the basis pursuit denosing (BPDN) algorithm \cite{BPDN} are chosen as the benchmark algorithms.
\begin{figure}[ht]
\begin{minipage}[t]{1\linewidth}
\centering
\includegraphics[width=3.2in]{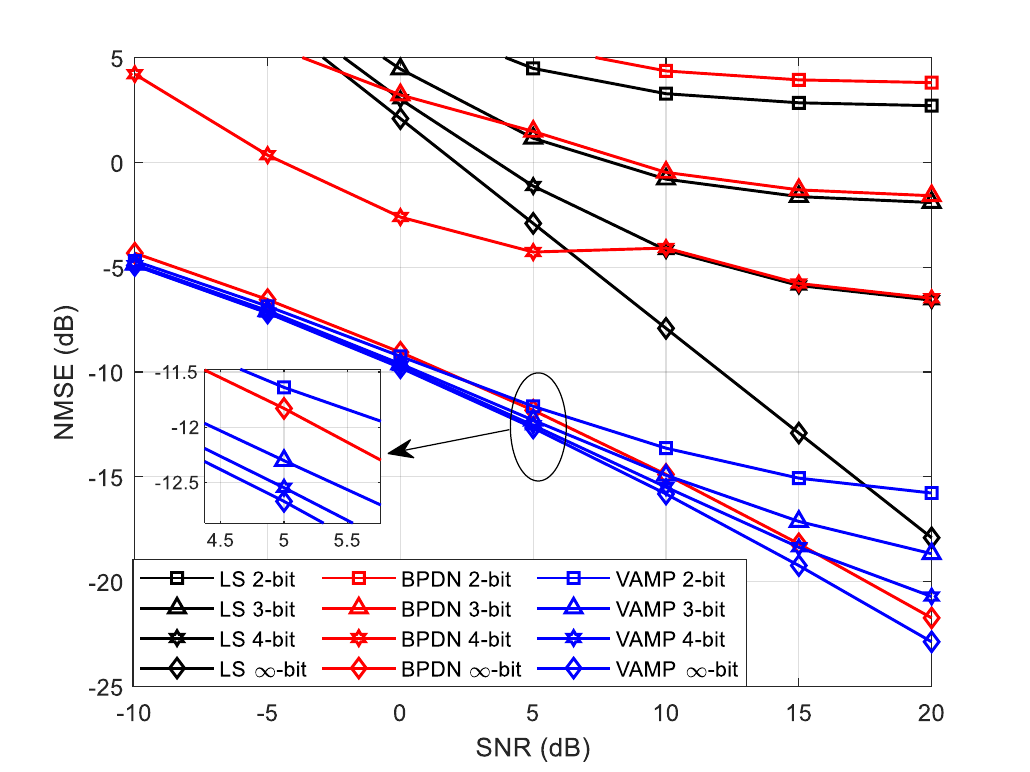}
\vspace{-0.2cm}
\caption{NMSE versus SNR, $K=4$.}
\label{fig5}\vspace{-0.5cm}
\end{minipage}%
\end{figure}

Fig.~\ref{fig5} illustrates the NMSE performance versus the SNR, where the number of users $K=4$, and the training length is set to $PT=512$. Compared with the LS algorithm and the BPDN algorithm, the Bayes inference based channel estimation method is more accurate. In addition, for low and medium SNR ($\leq$10 dB), the gap of NMSE between the few-bit ADCs (2-4 bits) and infinite-bit ADCs is less than 3 dB, which shows the applicability of few-bit ADCs in practice.

\begin{figure}[ht]
\begin{minipage}[t]{1\linewidth}
\centering
\includegraphics[width=3.2in]{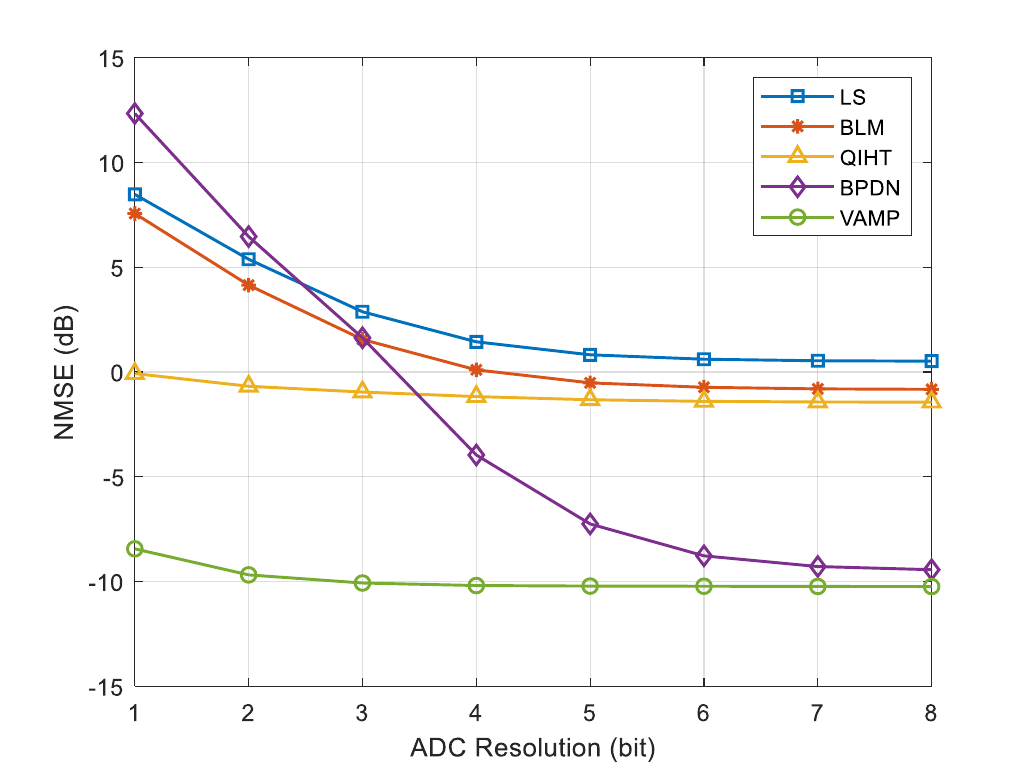}
\vspace{-0.2cm}
\caption{NMSE versus Bit number.}
\label{fig6}\vspace{-0.3cm}
\end{minipage}%
\hfill
\end{figure}

Fig.~\ref{fig6} shows the NMSE performance versus the number of bits for the multi-antenna user scenario, where SNR $=$ 0 dB. As shown in Fig.~\ref{fig6}, the proposed VAMP-based algorithms outperform the other benchmark algorithms. The BPDN algorithm performs similarly to VAMP when the ADC precision reaches to 6, 7, 8 and infinite bit, but performs worse for few-bit ADCs.

\begin{figure}[ht]
	\vspace{-0.2cm}
\begin{minipage}[t]{1\linewidth}
\centering
\includegraphics[width=3.2in]{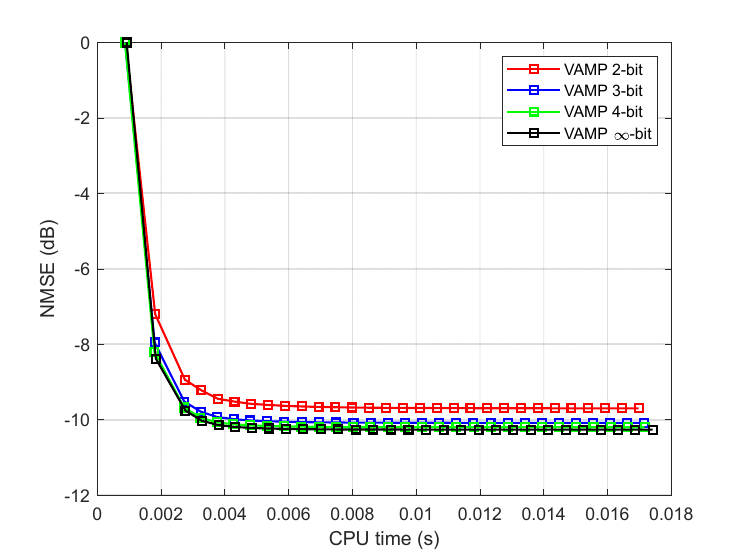}
\vspace{-0.2cm}
\caption{NMSE versus CPU time.}
\label{fig7}\vspace{-0.3cm}
\end{minipage}%
\hfill
\end{figure}

In Fig.~\ref{fig7}, the NMSE versus the runtime by varying the number of iterations for the VAMP algorithm under the multi-antenna user scenario is demonstrated, where SNR $=$ 0 dB. For the 2, 3 and 4-bit quantization, the VAMP based cascaded channel estimation algorithm only needs less than 10 iterations to converge and the runtimes are less than 0.02s due to the proposed FFT based fast implements, which demonstrates the low computational complexity for the VAMP based cascaded channel estimation algorithm.
\vspace{-0.2cm}

\section{Conclusion}\label{conclusion}
In this paper, we studied the cascaded channel estimation for an RIS-aided mmWave massive MIMO system, where the BS is equipped with few-bit ADCs and the BS, the RIS and the user are employed with UPAs. Considering the sparse property of the mmWave cascaded channel, we formulated the corresponding CS problem. To tackle with the information loss caused by coarse quantization, we adopted the Bayesian inference and VAMP algorithm to recovery the received signals and the sparse angular domain cascaded channel iteratively. The corresponding fast computation algorithm was derived to lower the complexity. The proposed strategy outperforms the other benchmark algorithms. Furthermore, the performance gap between low-precision ADCs and ideal infinite precision ADCs is acceptable for low and medium SNR, which implied that the low-precision ADCs is applicable for the low and medium SNR scenario.

\begin{appendix}
	\section{Proof of Proposition in \cite{junfang}}\label{derivationofcrb}
	We provide a detailed proof that ${\bf Y}=\left( {{\mathbf{S}}^{T}}\mathbf{F}_{Q}^{*}\otimes {{\mathbf{F}}_{N}} \right)\left( \mathbf{\widetilde{H}}_{\text{ru}}^{T}\otimes {{{\mathbf{\widetilde{H}}}}_{\text{br}}} \right)\left( \mathbf{F}_{M}^{T}\diamond \mathbf{F}_{M}^{H} \right)\mathbf{E}+ {\bf W}$ can be compressed as ${\bf Y}={\left( {{\mathbf{S}}^{T}}\mathbf{F}_{Q}^{*}\otimes {{\mathbf{F}}_{N}} \right){\bf \Lambda}\mathbf{F}_{M}^{H}\mathbf{E}}+{\bf W}$. Denotes $\mathbf{F}_M=\left[ \mathbf{f}_1,\cdots ,\mathbf{f}_M \right] $, then the Khatri-Rao product $\mathbf{F}_{M}^{T}\diamond \mathbf{F}_{M}^{H}$ can be written as
	\begin{equation}
		\left( \mathbf{F}_{M}^{T}\diamond \mathbf{F}_{M}^{H} \right) =\left[ \begin{array}{l}
			\mathbf{F}_{M}^{H}\mathrm{Diag}\left( \mathbf{f}_1 \right)\\
			\vdots\\
			\mathbf{F}_{M}^{H}\mathrm{Diag}\left( \mathbf{f}_M \right)\\
		\end{array} \right] .
	\end{equation}
	Since $\mathbf{F}_M=\mathbf{U}_{M_1}\otimes\mathbf{U}_{M_2}$, we have
	\begin{align}
		\mathbf{F}_{M}^{H}\mathrm{Diag}\left( \mathbf{f}_m \right) &=\left( \mathbf{U}_{M_1}^{H}\otimes \mathbf{U}_{M_2}^{H} \right) \mathrm{Diag}\left( \mathbf{f}_m \right) \nonumber
		\\
		&=\left( \mathbf{U}_{M_1}^{H}\otimes \mathbf{U}_{M_2}^{H} \right) \mathrm{Diag}\left( \left[ \begin{array}{c}
			u_{m_1,1}\mathbf{u}_{M_2}\left( :,m_2 \right)\\
			\vdots\\
			u_{m_1,M_1}\mathbf{u}_{M_2}\left( :,m_2 \right)\\
		\end{array} \right] \right) \nonumber\\
		&=\left( \begin{matrix}
			u_{1,1}^{\ast}u_{m_1,1}\mathbf{U}_{M_2}^{H}\mathrm{Diag}\left( \mathbf{u}_{M_2}\left( :,m_2 \right) \right)&		\cdots&		u_{M_1,1}^{\ast}u_{m_1,M_1}\mathbf{U}_{M_2}^{H}\mathrm{Diag}\left( \mathbf{u}_{M_2}\left( :,m_2 \right) \right)\\
			\vdots&		\ddots&		\vdots\\
			u_{1,M_1}^{\ast}u_{m_1,1}\mathbf{U}_{M_2}^{H}\mathrm{Diag}\left( \mathbf{u}_{M_2}\left( :,m_2 \right) \right)&		\cdots&		u_{M_1,M_1}^{\ast}u_{m_1,M_1}\mathbf{U}_{M_2}^{H}\mathrm{Diag}\left( \mathbf{u}_{M_2}\left( :,m_2 \right) \right)\\
		\end{matrix} \right) ,
	\end{align}
	where $m_1$ and $m_2$ are the corresponding entry and the column index of the $m$-th column of ${\bf F}_M=\mathbf{U}_{M_1}\otimes\mathbf{U}_{M_2}$, respectively. Note that the DFT matrix has the following property
	\begin{equation}
		\mathbf{U}_{M_2}^{H}\mathrm{Diag}\left( \mathbf{u}_{M_2}\left( :,m_2 \right) \right) =\mathbf{\Pi }_{m_2}\mathbf{U}_{M_2}^{H}.
	\end{equation}
	where the permutation matrix $\mathbf{\Pi }_{m_2}$ is associated with the index $m_2$. The reason of this property is that any two columns $\mathbf{u}_{M_2}\left( :,x_1 \right)$ and $\mathbf{u}_{M_2}\left( :,x_2 \right)$ from DFT matrix $\mathbf{U}_{M_2}$, the hadamard product of the two columns $\mathbf{u}_{M_2}\left( :,x_1 \right)\odot\mathbf{u}_{M_2}\left( :,x_2 \right)$ still belongs to the columns of the DFT matrix. Hence, we can prove that 
	\begin{align}
		&\left( \begin{matrix}
			u_{1,1}^{\ast}u_{m_1,1}\mathbf{U}_{M_2}^{H}\mathrm{Diag}\left( \mathbf{u}_{M_2}\left( :,m_2 \right) \right)&		\cdots&		u_{M_1,1}^{\ast}u_{m_1,M_1}\mathbf{U}_{M_2}^{H}\mathrm{Diag}\left( \mathbf{u}_{M_2}\left( :,m_2 \right) \right)\\
			\vdots&		\ddots&		\vdots\\
			u_{1,M_1}^{\ast}u_{m_1,1}\mathbf{U}_{M_2}^{H}\mathrm{Diag}\left( \mathbf{u}_{M_2}\left( :,m_2 \right) \right)&		\cdots&		u_{M_1,M_1}^{\ast}u_{m_1,M_1}\mathbf{U}_{M_2}^{H}\mathrm{Diag}\left( \mathbf{u}_{M_2}\left( :,m_2 \right) \right)\\
		\end{matrix} \right)\nonumber\\
		&\qquad=\left( \begin{matrix}
			u_{1,1}^{\ast}u_{m_1,1}\mathbf{\Pi }_{m_2}\mathbf{U}_{M_2}^{H}&		\cdots&		u_{M_1,1}^{\ast}u_{m_1,M_1}\mathbf{\Pi }_{m_2}\mathbf{U}_{M_2}^{H}\\
			\vdots&		\ddots&		\vdots\\
			u_{1,M_1}^{\ast}u_{m_1,1}\mathbf{\Pi }_{m_2}\mathbf{U}_{M_2}^{H}&		\cdots&		u_{M_1,M_1}^{\ast}u_{m_1,M_1}\mathbf{\Pi }_{m_2}\mathbf{U}_{M_2}^{H}\\
		\end{matrix} \right) \nonumber\\
		&\qquad=\left( \begin{matrix}
			u_{m_1,1}\mathbf{\Pi }_{m_2}\mathbf{U}_{M_2}^{H}&		\cdots&		u_{m_1,M_1}\mathbf{\Pi }_{m_2}\mathbf{U}_{M_2}^{H}\\
			\vdots&		\ddots&		\vdots\\
			u_{m_1+1,1}\mathbf{\Pi }_{m_2}\mathbf{U}_{M_2}^{H}&		\cdots&		u_{m_1+1,M_1}\mathbf{\Pi }_{m_2}\mathbf{U}_{M_2}^{H}\\
		\end{matrix} \right) \nonumber\\
		&\qquad=\left( \mathbf{\Pi }_{m_1}\mathbf{U}_{M_1}^{H} \right) \otimes \left( \mathbf{\Pi }_{m_2}\mathbf{U}_{M_2}^{H} \right) .
	\end{align}
	Analogously, it can be derived that the last $M(M-1)$ rows are the permutation version of the first $M$ rows and they still belongs to the first $M$ rows. Based on the above discussion, we can obtain that ${\bf \Lambda}$ is the compressed version of $\left( \mathbf{\widetilde{H}}_{\text{ru}}^{T}\otimes {{{\mathbf{\widetilde{H}}}}_{\text{br}}} \right)$
	\begin{equation}
		\mathbf{\Lambda }\left( :,i \right) =\sum_{n\in \mathcal{S} _i}{\left( \widetilde{\mathbf{H}}_{\mathrm{ru}}^{T}\otimes \widetilde{\mathbf{H}}_{\mathrm{br}} \right) _{:,n}},
	\end{equation}
	where $\mathbf{\Lambda }\left( :,i \right)$ denotes the $i$-th column of $\bf \Lambda$ and $\mathcal{S} _i$ denotes the set of indices associated with those rows in $\left( \mathbf{F}_{M}^{T}\diamond \mathbf{F}_{M}^{H} \right)$ that are identical to the $i$-th row of $\left( \mathbf{F}_{M}^{T}\diamond \mathbf{F}_{M}^{H} \right)$ \cite{junfang}.
\end{appendix}

\bibliographystyle{IEEEtran}
% %argument is your BibTeX string definitions and bibliography database(s)
\bibliography{myre}

% that's all folks

\end{document}